\documentclass{emulateapj}

\shorttitle{H2$^{*}$ near Herschel 36}
\shortauthors{Rachford et al.}

\begin{document}

\title{Vibrationally Excited Molecular Hydrogen Near Herschel 36}

\author{Brian L. Rachford}
\affil{Department of Physics, Embry-Riddle Aeronautical University,
3700 Willow Creek Road, Prescott, AZ 86301-3720; rachf7ac@erau.edu}

\author{Theodore P. Snow}
\affil{Center for Astrophysics and Space Astronomy, Department of 
Astrophysical and Planetary Sciences, University of Colorado, Boulder, CO
80309-0389}

\author{Teresa L. Ross}
\affil{Department of Astronomy, New Mexico State University, Las Cruces, NM 
88003-8001}

\begin{abstract}
We present the first high resolution UV spectra toward Herschel 36, a
Trapezium-like system of high-mass stars contained within the Lagoon Nebula
(M8, NGC 6523).  The spectra reveal extreme ro-vibrational excitation of
molecular hydrogen in material at a single velocity or very small range of
velocities, with this component presumably lying near the star system and
undergoing fluorescent excitation.  The overall H$_2$ excitation is similar to,
but apparently larger than, that seen towards HD 37903 which previously showed
the largest vibrationally excited H$_2$ column densities seen in UV absorption
spectra.  While the velocities of the highly excited H$_2$ lines are consistent
within each observation, it appears that they underwent a $\sim$60 km s$^{-1}$
redshift during the 3.6 years between observations.  In neither case does the
velocity of the highly excited material match the velocity of the bulk of the
line-of-sight material which appears to mostly be in the foreground of M8.
Recent work shows unusually excited CH and CH$^{+}$ lines and several unusually
broad Diffuse Interstellar Bands towards Herschel 36.  Along with the
H$_2$ excitation, all of these findings appear to be related to the extreme
environment within $\sim$0.1 pc of the massive young stellar system.
\end{abstract}


\section{Introduction}
The Lagoon Nebula (M8, NGC 6523) is one of the most prominent bright nebulae in
the sky, and contains regions of recent star formation and many early-type
stars.  The optically brightest portion of the nebula lies near the object
Herschel 36\footnote{This star is sometimes associated with the designation HD
164740, but the relevant entry in the Henry Draper catalog (Cannon \& Pickering
1922) indicates that object 164740 is nebular.  This is presumably the bright
Hourglass region described in the next paragraph, and not the star described by
J. Herschel (1847), which has the identifier CD -24 13806.}.  Previously
considered a single O-type star, Herschel 36 has recently been
spectroscopically resolved into a O7.5 V star in a $\sim$500-day, $\sim$3 au 
mutual orbit with a close late-O/early-B binary (Arias et al.\ 2010).  The
presence of an additional heavily embedded star of early B-type has been
inferred 0.25 arcsec (375 au at 1500 pc) southeast of the triplet (Goto et al.\
2006).  The Herschel 36 system thus has much in common with the Trapezium
system in the Orion Nebula.

The hot stars, presumably dominated by the O7.5 V component, are the primary
illuminating source for the bright region 15$\arcsec$ east of Herschel 36,
known as the Hourglass for its distinctive shape in optical images (Thackeray
1950, Wolff 1961).  Additional hot stars ionize the extended \ion{H}{2} region
surrounding the Hourglass.  Infrared radiation from the embedded early B-type
star may be responsible for the unusually high excitation of CH and CH$^{+}$
seen in optical spectra towards Herschel 36 (Oka et al.\ 2013).

The Herschel 36 system is highly obscured by dust (E(B-V) = 0.87), and the line
of sight shows exceptional extinction characteristics.  The total-to-selective
extinction ratio, $R_V$ $\equiv$ $A_V$/$E(B-V)$ = 5.21 $\pm$ 0.10 (Fitzpatrick
\& Massa 2007), is one of the largest known and a correction for foreground
material suggests an even larger $R_V$ for material local to Herschel 36 (Hecht
et al.\ 1982).  This value is consistent with a population of larger than
normal dust grains.  The far-UV extinction curve is correspondingly shallow
(Fitzpatrick \& Massa 2007), and it is the very weak UV extinction that allows
sufficient transmission of light for far-UV observations.

While the Hourglass and the general region around Herschel 36 have been studied
extensively at various wavelengths (see Dahlstrom et al.\ 2013 for a review)
there has been no previous UV spectroscopy toward Herschel 36 with sufficient
resolution and S/N for interstellar absorption line studies.  In this paper, we
give the results of such an investigation, focusing on the highly unusual
molecular hydrogen absorption seen along the line of sight.

\section{Overview of the data}
\subsection{Observations}
Herschel 36 was observed twice by the {\it Far Ultraviolet Spectroscopic
Explorer} ({\it FUSE}) as part of the PI Team ``translucent'' cloud program
(Rachford et al.\ 2002); see Table 1 for observation details.  Although our
initial analysis involved earlier data products, for our final results we used
data downloaded from the Mikulski Archive for Space Telescopes (MAST) in late
2011, which was processed with version 3.2.1 of the CALFUSE pipeline.  While
data are present shortward of 1000 \AA\ down to near the Lyman limit, the S/N
is quite poor at these wavelengths.  Thus, we have focused on data longward of
1000 \AA\ in the LiF 1A channel (987.4--1082.7 \AA ) and the LiF 2A channel
(1086.9--1181.5 \AA ) (see Moos et al.\ 2000 for more information on the
configuration of the {\it FUSE} spectrograph).  The velocity resolution is
$\Delta$$v$ $\approx$ 15--20 km/s, depending on wavelength and instrument
channel.  Outside of the deep absorption bands of H$_2$ and other strong
absorption features, the S/N per resolution element is typically 15--30 with
the best S/N occurring at wavelengths longer than 1100 \AA .

\begin{deluxetable}{cccc}
\tablecaption{{\it FUSE} Observations}
\tablewidth{0pt}
\tablenum{1}
\tablehead{
  \colhead{Data Set ID} & \colhead{Date} & \colhead{Exp.\ time} & \colhead{Aperture} \\
  & & \colhead{(ksec)}
}
\startdata
P1162001 & 2000 Aug 30 &    10.6 & LWRS \\
P1162002 & 2004 Apr 8  & \phn5.9 & LWRS
\enddata
\end{deluxetable}

\subsection{The H$_2$ spectrum}
The overall far-UV spectrum is typical of significantly reddened stars, as it
is dominated by absorption from \ion{H}{1} and H$_2$, the latter in the form of
damped ro-vibrational bandheads due to absorption from the $J$ = 0 and $J$ = 1
levels.\footnote{Throughout this paper we will apply the usual convention of
writing the lower rotational level of an absorption transition, $J''$, as
simply $J$, and similarly writing the lower vibrational level as $v$.}
However, a closer look at the spectra reveals three unusual features.

First, we see a large number of weak lines corresponding to absorption from
excited vibrational states.  A sample of these lines is given in Figure 1.
Vibrational excitation of H$_2$ in diffuse and translucent lines of sight of
sufficient magnitude to be detected with {\it FUSE} is quite rare.  Prominent
examples include the lines of sight toward HD 34078 (Boiss\'{e} et al.\ 2005),
and HD 37903 (Gnaci\'{n}ski 2011).  Only the latter is comparable with the Herschel
36 spectrum in terms of the quantity and strength of vibrationally excited
lines.  Second, when looking at lines from the $J$ = 2--4 levels of the ground
vibrational state, we see a broad velocity structure spread across tens of km
s$^{-1}$.  Finally, we see a significant velocity shift in a portion --- but
not all --- of the material between the first observation and the second.

\begin{figure*}
\plotone{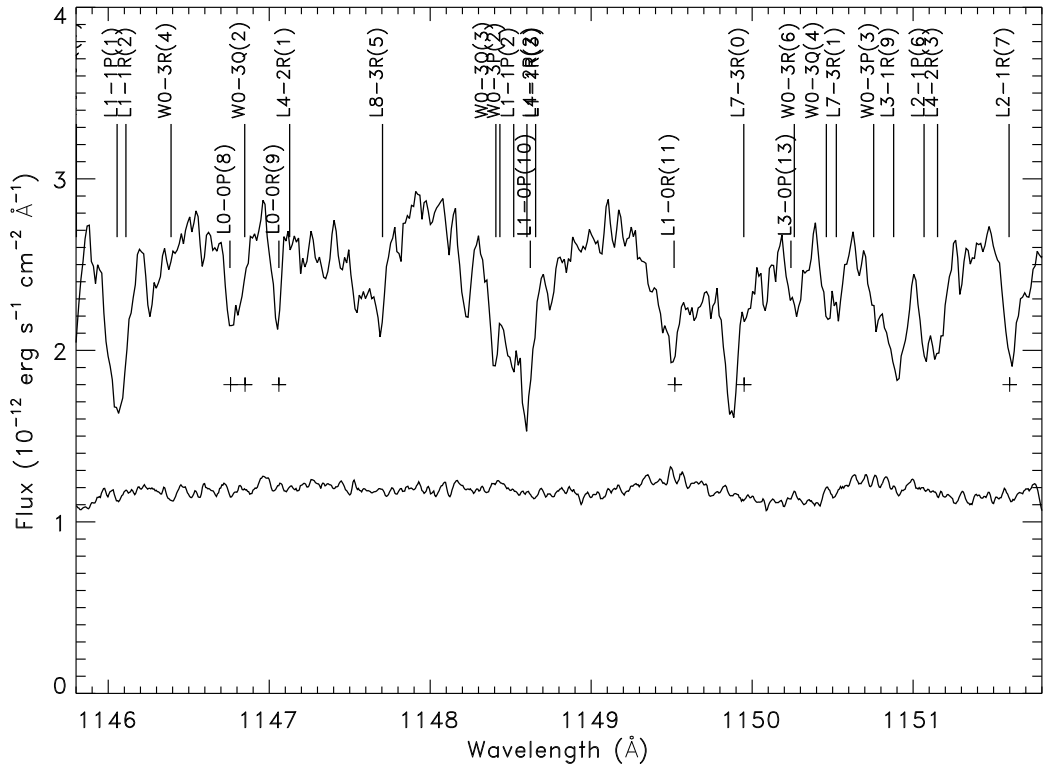}
\caption{A portion of the second observation showing numerous lines from
vibrationally excited states (upper row of labels and long tick marks), and
high-excitation states of the vibrational ground state.  The lower rotational
level of the transition ($J$) is given in parentheses and the lower vibrational
level ($v$) is given before the branch label ($P$, $Q$, or $R$).  The {\it FUSE}
spectrum of M8 star HD 164906 is shown for comparison, with the continuum
normalized to about half of the Herschel 36 continuum.  The lack of ISM
features in this portion of the spectrum is the overwhelmingly typical case in
{\it FUSE} spectra.  Weak line identifications for the Herschel 36 spectrum are
generally not included, nor are atomic lines.  Lines specifically used in later
analysis are indicated directly below the line.}
\end{figure*}

The highly excited material is a compelling issue on its own; in fact it
appears to be the strongest such excitation seen to date in ultraviolet
absorption spectra.  Recent work has also revealed unusually strong excitation
of the CH and CH$^{+}$ molecules along this line of sight, as well as extended
redward wings in several Diffuse Interstellar Bands (DIBs) consistent with
excitation of higher than normal rotational states in the molecules that
presumably produce the DIB absorption (Dahlstrom et al.\ 2013)

A full understanding of the velocity structure and changes in velocity structure
seen in the {\it FUSE} spectra will likely require additional data and
analysis.  However, considerable information on excited H$_2$ is available from
just the existing {\it FUSE} spectra.  Thus, in this paper we will mainly focus
on the excited H$_2$ and explore the other two issues only as they directly
relate to the highly excited material.

\section{Analysis and interpretation}
\subsection{Line measurements and curve-of-growth analysis}
The lines from $J$ $\geq$ 5 of the ground vibrational state, and lines from the
excited vibrational states are narrow enough to be consistent with the
resolution of the spectrograph.  Thus, any velocity structure is buried within
the resolution element, and we will treat those lines as representing a single
component that we call the ``highly excited'' component.  It is probable
that the strongest $J$ = 5 lines show a weak wing that represents a small
fraction of the absorption.  However, our analysis primarily relies upon the
weakest $J$ = 5 lines.  As a predictive guide to the vibrationally excited
lines, we estimated the strength of each putative line based on column
densities toward HD 37903 derived from a high resolution {\it HST STIS}
spectrum (Meyer et al.\ 2002) which shows similar H$_2$ excitation to our
spectrum of Herschel 36.  We then selected lines from the Abgrall et al.\
(1993a,b) lists that were most likely to be unblended and visible at the lower
resolution and sensitivity of the {\it FUSE} spectra.

Given that the observed excited H$_2$ line profiles are dominated by the {\it
FUSE} instrumental profile, which is reasonably modeled by a Gaussian
especially at low to moderate S/N (e.g., Jensen et al.\ 2010), we measured lines
by fitting Gaussian profiles.  Thus, we obtained central wavelength, line depth,
and line width, the latter two quantities directly give the equivalent width of
the line and a formal uncertainty.  We combined this uncertainty in quadrature
with a continuum uncertainty estimated from the line width and the noise in the
continuum fit.  In many cases, this likely overestimates the true
uncertainties.  However, continuum placement is a particular challenge here
given the likelihood of otherwise undetected vibrationally excited lines
contaminating the continuum, analogous to the ``line fog" seen in optical
spectra of late-type stars.  When possible, adjacent H$_2$ lines were deblended
with multiple Gaussians.

Although our technique was to measure lines and perform a curve-of-growth
analysis, there is significant line confusion due to the wealth of detectable
transitions, most of which have never been observed in {\it FUSE} spectra.
Thus, we followed an iterative process whereby we generated a model spectrum from
the curve-of-growth analysis, used that to check if the levels were reasonably
modeled, and then explored individual lines again.  Many lines were eliminated
based on the modeling suggesting that a stronger line may be contaminated by a
weaker line.  This procedure leverages the strength of the curve-of-growth
method (i.e., deriving the $b$-value and column densities simultaneously with
the cleanest lines) while not having to fully model the continuum, stellar
flux, and instrumental function across the entire spectrum.

Despite the first observation having a longer exposure time, the second
observation was of higher quality and we were able to measure more lines.  Thus
we used the second observation as the primary source for the excited lines.
Unfortunately, even in the second observation most lines were detected below
the 3-$\sigma$ level relative to the combined measurement uncertainties and
continuum uncertainties.

Our final list included 122 lines from 37 ro-vibrational levels in the second
observation, and we successfully measured 72 of those lines in the first
observation.  Figure 2 shows that there is reasonable agreement in the
equivalent widths of these lines, albeit with large uncertainties.  This
indicates that there was not a dramatic change in the strength of the excited
component between the two observations, and allows us to treat the second
observation as representative.  We note that absorption from additional
ro-vibrational levels is evident in the second spectrum, but we limited the
analysis to levels where we were most confident that at least one
uncontaminated line could be decisively measured.

\begin{figure}
\plotone{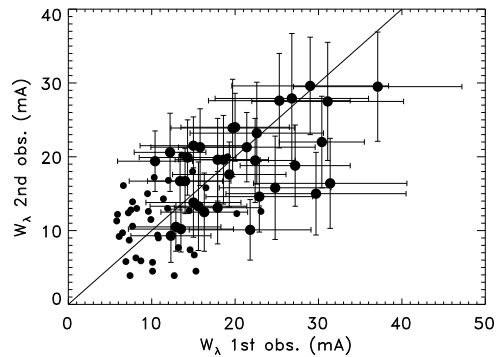}
\caption{Equivalent width comparison between observations.  Lines detected at
above the 2-sigma level are shown with error bars and larger symbols.  Solid
line: unit slope passing through zero.}
\end{figure}

\begin{figure}
\plotone{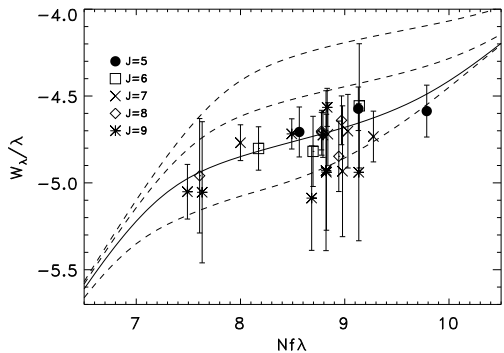}
\caption{Best-fit curve of growth ($b$=1.2 km s$^{-1}$) for second observation,
with lines from the $J$ = 5--9 levels from the ground vibrational state, which
provide most of the weight in the fit.  Curves of growth for $b$ = 2.4 and 4.8
km s$^{-1}$ are shown above the best fit, and $b$ = 0.6 km s$^{-1}$ is shown
below.}
\end{figure}

We performed least squares fits of the equivalent widths from the second
observation to a grid of single-component curves of growth with 0.1 km s$^{-1}$
spacing in $b$.  We attempted fits for lines from various combinations of
ro-vibrational states and consistently found small $b$-values less than 2 km
s$^{-1}$ for most of these combinations.  Thus, we adopted the overall best-fit
of $b$ = 1.2 km s$^{-1}$.  In Figure 3, we show the $J$ = 5--9 levels of the
vibrational ground state, which carry most of the weight in the curve of growth
fits.  The resulting column densities for all measured levels are given in
Table 2.

\begin{deluxetable*}{cccccc}
\tablecolumns{3}
\tablewidth{0pc}
\tablenum{2}
\tablecaption{Logarithmic H$_2$ column densities for the highly excited material, 2nd observation\tablenotemark{a}}
\tablehead{$J$ &$N(v=0)$  &$N(v=1)$  &$N(v=2)$	&$N(v=3)$  &$N(v=4)$}

\startdata
0  &                            &14.54$^{+0.78}_{-0.57}$ (4) &13.49$^{+3.75}_{-1.30}$ (2) &13.71$^{+4.66}_{-1.74}$ (1) &13.88$^{+1.01}_{-0.67}$ (2) \\
1  &                            &14.25$^{+1.28}_{-0.59}$ (3) &14.17$^{+1.31}_{-0.62}$ (3) &13.74$^{+0.72}_{-0.36}$ (3) &13.38$^{+1.03}_{-0.47}$ (2) \\
2  &                            &15.22$^{+2.09}_{-0.94}$ (2) &14.04$^{+1.35}_{-0.62}$ (1) &13.84$^{+1.22}_{-0.49}$ (3) &13.33$^{+0.54}_{-0.32}$ (2) \\
3  &                            &15.26$^{+0.87}_{-0.87}$ (9) &14.36$^{+1.57}_{-0.62}$ (1) &14.19$^{+2.55}_{-1.20}$ (3) &13.93$^{+1.63}_{-0.62}$ (1) \\
4  &                            &15.16$^{+0.80}_{-0.81}$ (5) &14.14$^{+1.16}_{-0.53}$ (2) &14.45$^{+1.72}_{-0.65}$ (5) &  \\
5  &16.66$^{+0.53}_{-0.72}$ (3) &14.73$^{+1.10}_{-0.71}$ (6) &13.87$^{+0.98}_{-0.41}$ (1) &13.45$^{+0.78}_{-0.42}$ (2) &  \\
6  &15.71$^{+0.97}_{-0.81}$ (3) &14.32$^{+1.73}_{-0.67}$ (4) &                            &  &  \\
7  &16.08$^{+0.70}_{-0.66}$ (5) &15.13$^{+0.99}_{-0.94}$ (4) &14.25$^{+1.54}_{-0.57}$ (1) &13.55$^{+1.31}_{-0.50}$ (2) &  \\
8  &15.70$^{+0.80}_{-0.87}$ (6) &                            &13.51$^{+0.56}_{-0.34}$ (1) &  &  \\
9  &15.74$^{+0.93}_{-0.74}$ (11) &                           &                            &13.69$^{+1.07}_{-0.47}$ (2) &  \\
10 &15.05$^{+1.33}_{-0.73}$ (5) &  &  &  &  \\
11 &14.93$^{+1.09}_{-0.59}$ (5) &                            &                            &  &  \\
12 &14.19$^{+2.12}_{-0.71}$ (2) &  &  &  &  \\
13 &14.48$^{+1.54}_{-0.61}$ (5) &                            &                            &  & 
\enddata
\tablenotetext{a}{The number in parentheses behind the column density is the number
of lines measured in that level.}
\end{deluxetable*}

Our $b$-value and column densities are rather poorly constrained due to the
large relative errors on most individual lines.  Based on the curve-of-growth
analysis for the measured lines, the 1$\sigma$ error range for the $b$-value is
0.1--4.1 km s$^{-1}$.  However, we have the additional constraint of modeling
the spectra which allows us to look at blended lines and undetected lines in
addition to the relatively small fraction of isolated lines.  Thus, we are more
confident in the derived $b$-value than would be warranted by purely the
analysis of measured lines.

The derived $b$-value for the highly excited H$_2$ is quite small; we are
unaware of any smaller values ever found in absorption.  Other lines of sight
with considerable vibrational excitation have also shown relatively small
$b$-values with a single component even at a resolution of $\sim$3 km s$^{-1}$
from {\it HST} data (Meyer at al. 2001 and Gnaci\'{n}ski 2011 for HD 37903 from
{\it FUSE} and {\it HST}; Boiss\'{e} et al 2005 for HD 34078 from {\it FUSE};
Gnaci\'{n}ski 2013 for HD 147888 {\it FUSE} and {\it HST}).  Smaller $b$-values
of less than 1 km s$^{-1}$ are seen for individual components of various atomic
species in many lines of sight in spectra with $\sim$1 km s$^{-1}$ resolution
(e.g., Welty \& Hobbs 2001).  However, due to the lack of ultra high resolution
UV observations, we do not know if this holds for H$_2$ as well, and the low
mass of the molecule also works against very small $b$-values.  For purely
thermal broadening, $b_{\rm H2}$ = 1.2 km s$^{-1}$ corresponds to a temperature
of 174 K.  This should be contrasted with the $E/k$ values for the $v$ = 1
rotational levels which are $\gtrsim$6000 K, rising to $\gtrsim$22000 K for $v$
= 4, indicating that the observed excitation is due to a non-thermal process.

\begin{figure}
\plotone{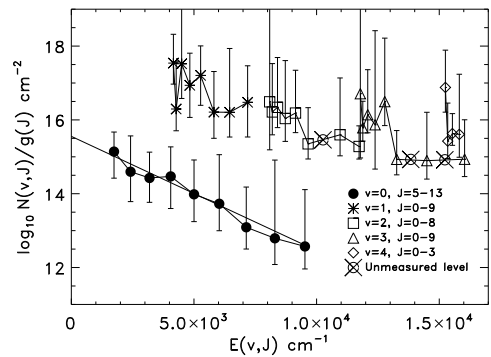}
\caption{Excitation diagram for the second observation of Herschel 36.  The
excited vibrational levels have each been shifted upward by three dex relative
to the $v$ = 0 level for clarity.  Unmeasured levels are labeled for clarity at the
interpolated location between adjacent levels.  The Boltzmann excitation fit for
$J$ = 5--9 of the $v$ = 0 level is shown, and extrapolated to cover $J$ =
0--13.}
\end{figure}

To further assess the excitation, in Figure 4, we present an excitation diagram
based on the column densities in Table 2.  The large uncertainties are
particularly apparent here, but the expected decreasing trend with increasing
$J$ of the quantity column density divided by statistical weight is generally
seen.  The uncertainties make a more detailed quantitative modeling analysis
difficult, as a broad range of cloud models could fit the column densities
within the large uncertainties.  However, a comparison with the apparently
similar and better constrained situation with HD 37903 may be instructive.

HD 37903 was analyzed by Meyer et al.\ (2001) using only the high-quality {\it
STIS} data, and has been recently re-analyzed by also including {\it FUSE} data
which provided column densities for non-vibrationally excited levels
(Gnaci\'{n}ski 2011).  In both analyses, interstellar cloud models favored an
interpretation whereby the extreme H$_2$ excitation was occurring in a
relatively dense cloud of material $\sim$0.5 pc away from HD 37903.

\begin{figure*}
\plotone{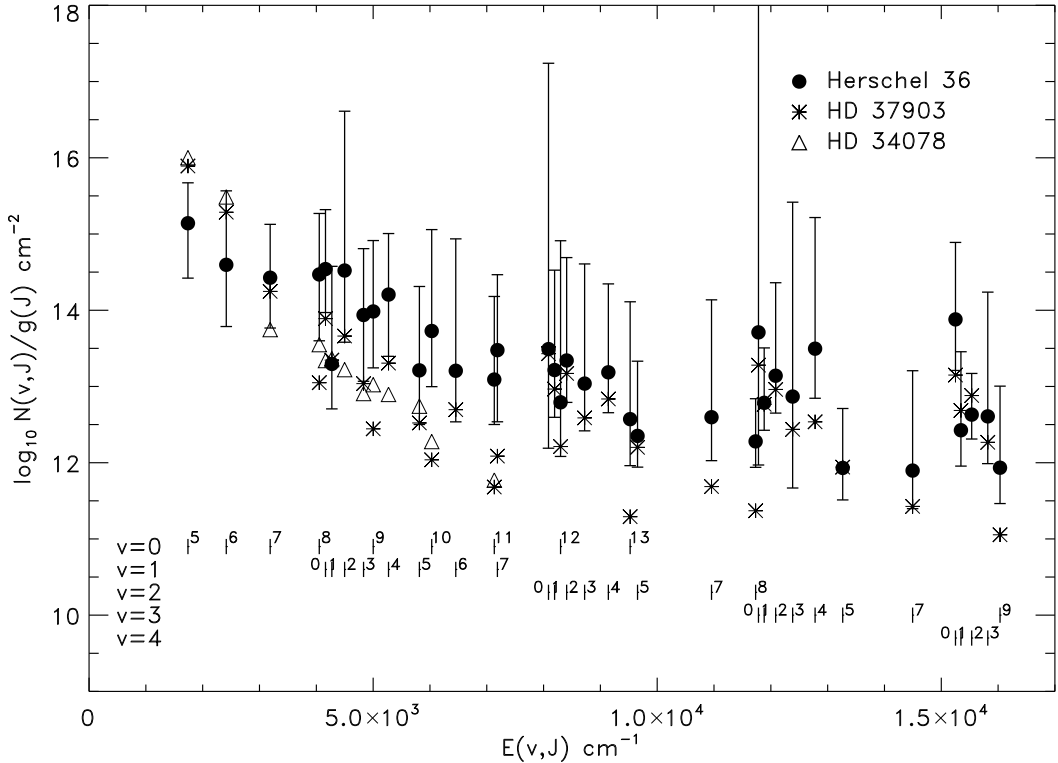}
\caption{Excitation diagram comparison for Herschel 36, HD 37903, and HD 34078.
Level identifications are given at the bottom of the plot with $v$ at the left,
and $J$ near the appropriate tick mark.}
\end{figure*}

In Figure 5, we plot the excitation diagram for Herschel 36, along with those
for HD 37903 (Gnaci\'{n}ski 2011) and HD 34078 (Boiss\'{e} et al 2005).  These
are the two lines of sight that show the strongest vibrationally excited lines,
although HD 34078 only has column densities for relatively few highly excited
levels.  Again, our values are highly uncertain, but the three lines of sight
all show relatively large amounts of excited material, with evidence that the
Herschel 36 column densities are even larger than those towards the other two
stars.  These lines of sight should be contrasted with the finding of Jensen et
al.\ (2010) of 3$\sigma$ upper limits of $\lesssim$10$^{13}$ cm$^{-2}$ for the
$v$ = 1, $J$ = 0 level for nearly all {\it FUSE} translucent lines of sight as
compared with our logarithmic column density of 14.54 for Herschel 36, 13.89
for HD 37903, and 13.34 for HD 34078.  This means that vibrationally excited
column densities for most lines of sight would be well below those seen in
Figure 5.  The few other lines of sight where vibrationally excited H$_2$ has
been detected in absorption also show smaller column densities those in Figure
5; i.e., Federman et al.\ (1995) for $\zeta$ Oph, Jensen et al.\ (2010) for HD
38087 and HD 199579, and Gnaci\'{n}ski (2013) for HD 147888.  In none of the
cases has velocity structure been seen within vibrationally excited lines, even
in {\it HST} data with 3 km s$^{-1}$ resolution.

\begin{figure*}
\plotone{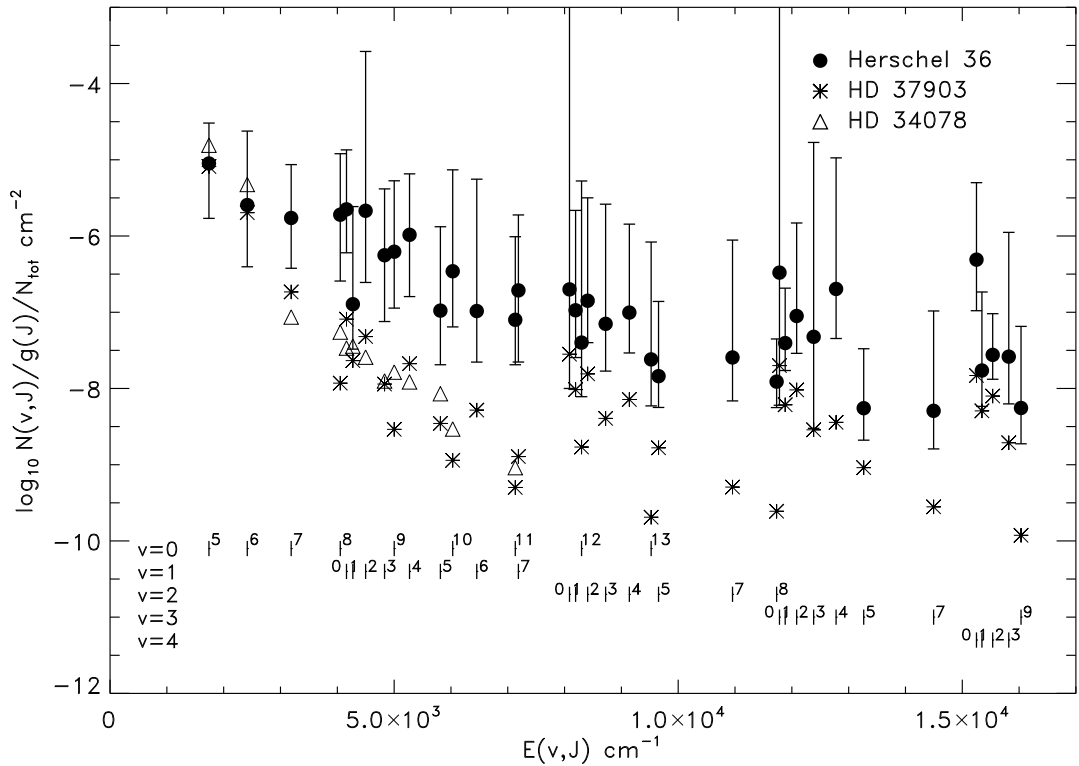}
\caption{Excitation diagram comparison for Herschel 36, HD 37903, and HD 34078,
normalized to the total H$_2$ column density for the line of sight.  Otherwise
the same as Fig. 5.}
\end{figure*}

For a different perspective, in Figure 6, we plot the column densities of the
excited levels relative to the total line-of-sight H$_2$ column densities.
Since the total H$_2$ column density towards Herschel 36 is smaller than for
the other two lines of sight, the differences are exaggerated.  However, in
reality, the differences may be even greater.  As we discuss in \S\ 3.3, much
of the line-of-sight material toward Herschel 36 may lie in the foreground of
M8 and not be directly related to the highly excited material near the star
itself.  Thus, the relative column densities of the highly excited levels may
be even higher in the material that lies within M8 itself.

\subsection{The highly excited component in the context of the line of
sight material}
Figures 7 and 8 reveal the difficulty in interpreting low-excitation states in
the highly excited component.  Broad structure appears for the $J$ = 2--4 lines
(we omit $J$ = 4 because with the weaker lines, the structure is not as
obvious), and the highly excited component does not match either of the
apparent peaks of the $J$ = 2--3 structure.  Nor does the highly excited
component appear to represent more than a small fraction of the observed
velocity range of $J$ = 2--3.  If the highly excited component were to actually
represent either of the $J$ = 2--3 peaks or the bulk of the $J$ = 0--1
profiles, it would require an unrealistic velocity shift of more than 20 km
s$^{-1}$ between $J$ = 5 and the lower levels.

\begin{figure}
\plotone{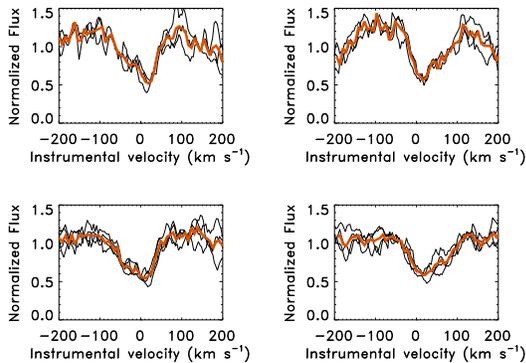}
\caption{Broad velocity structure in several $J$ = 2 and $J$ = 3 lines in the
first and second observations.  Upper left: $J$ = 2, first observation; Upper
right: $J$ = 2, second observation; Lower left: $J$ = 3, first observation;
Lower right: $J$ = 3, second observation.  Three lines of similar strength are
shown for $J$ = 2, and four lines are shown for $J$ = 3.  The thick lines are
averages.}
\end{figure}

\begin{figure*}
\plotone{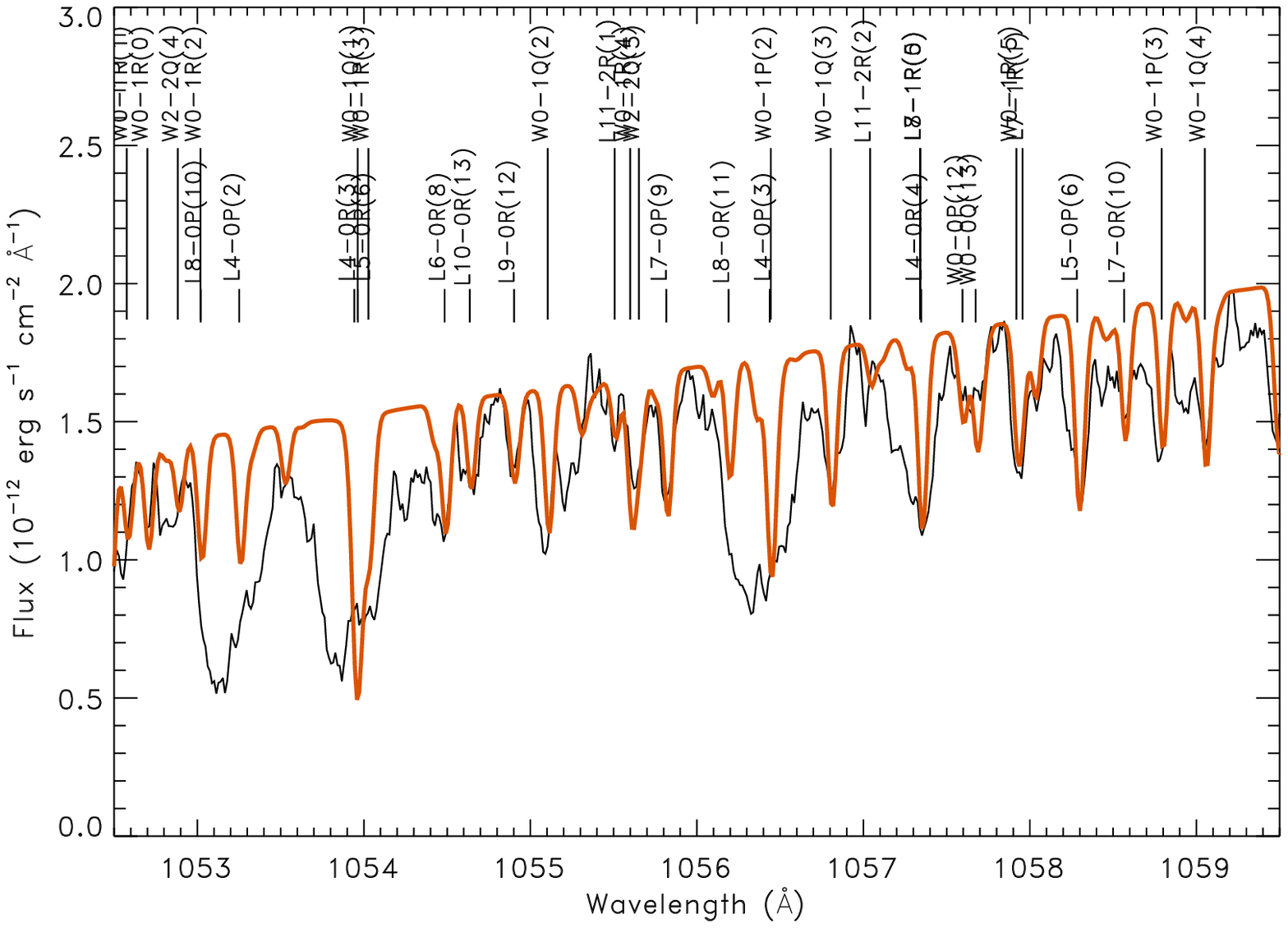}
\caption{Model of the highly excited component compared with second
observation.  Note that the model continuum is only approximate.  Atomic lines
(e.g., \ion{Fe}{2} $\lambda$1055) have not been modeled and the $J$ = 2--4
column densities for the highly excited component have been estimated using the
excitation diagram in Figure 4 as described in the text.  For clarity, only the
strongest vibrationally excited lines are labeled.}
\end{figure*}

In modeling the spectrum of the highly excited component for a variety of
ro-vibrational bandheads and thus a range of oscillator strengths, we find that
$J$ = 0--1 logarithmic column densities $\gtrsim$18.5 clearly ``overshoot'' the
observed profiles at those wavelengths.  These values are much smaller than the
total line of sight values reported by Rachford (2009); 19.92 and 19.86,
respectively.

We can set an uncertain {\it lower} limit on the low-$J$ column densities of
the highly excited material by using the excitation diagram in Figure 4.  If we
fit a line to the $J$ = 5--10 levels of the ground vibrational state, we obtain
a Boltzmann excitation temperature of 2019$^{+5675}_{-875}$ K (a fit of $J$ = 5--8
gives a similar result).  If we extrapolate this line to lower levels, we
obtain logarithmic column densities of 15.48, 16.40, 16.08, 16.60, 16.10, for
$J$ = 0, 1, 2, 3, and 4, respectively.  While the column densities for $J$ = 0
and $J$ = 1 may be significantly underestimated by this fit, it is common for
$J$ = 2 or 3 to 6 and beyond to be reasonably fit by a single line (see, e.g.,
Jensen et al.\ 2010).  Unless we dramatically increase the $b$-values, column
densities much greater than these for $J$ = 2--4 will ``overshoot'' the
observed profiles.  The evidence for $b$-value changes as a function of $J$
(e.g. Jenkins \& Pembert 1997; Lacour et al. 2005) is that the more highly
excited levels that may show larger $b$-values, a trend we do not see in our
data.  Significantly smaller column densities are not physically compatible
with the levels that we have been able to measure; i.e., they would not follow
a Boltzmann equilibrium trend.

Based on the arguments in the previous two paragraphs, our best estimates of
the logarithmic column densities of the $J$ = 0--4 levels of the highly excited
component are $\lesssim$18.0, $\lesssim$18.0, 16.1, 16.6, and 16.1,
respectively.  While the material responsible for the extended velocity
structure may lie within M8, it does not produce measurable absorption in the
more highly excited levels above $J$ = 5, and thus must be much further from a
strong UV source, and/or strongly shielded from UV radiation.

The final difficulty in interpreting the highly excited component is a
significant velocity shift during the 3.4 year gap between observations.  Figure
9 shows both observations with the wavelength solutions given by the CALFUSE
3.2.1 pipeline.  For clarity, only lines from the vibrational ground state have
been labeled; recall that levels $J$ $\geq$ 5 are part of the highly excited
material that shows no component structure.  The line labels have not been
shifted in this plot, and they line up to within several km s$^{-1}$ of the
appropriate lines in the spectrum, as do the excited lines themselves.
However, the $J$ = 0--1 bandhead is clearly offset between the two
observations.  In contrast, as shown in Figure 10, if we shift the second
observation redward by 60 km s$^{-1}$, the $J$ = 0--1 bandheads line up, while
the excited material does not.

\begin{figure*}
\plotone{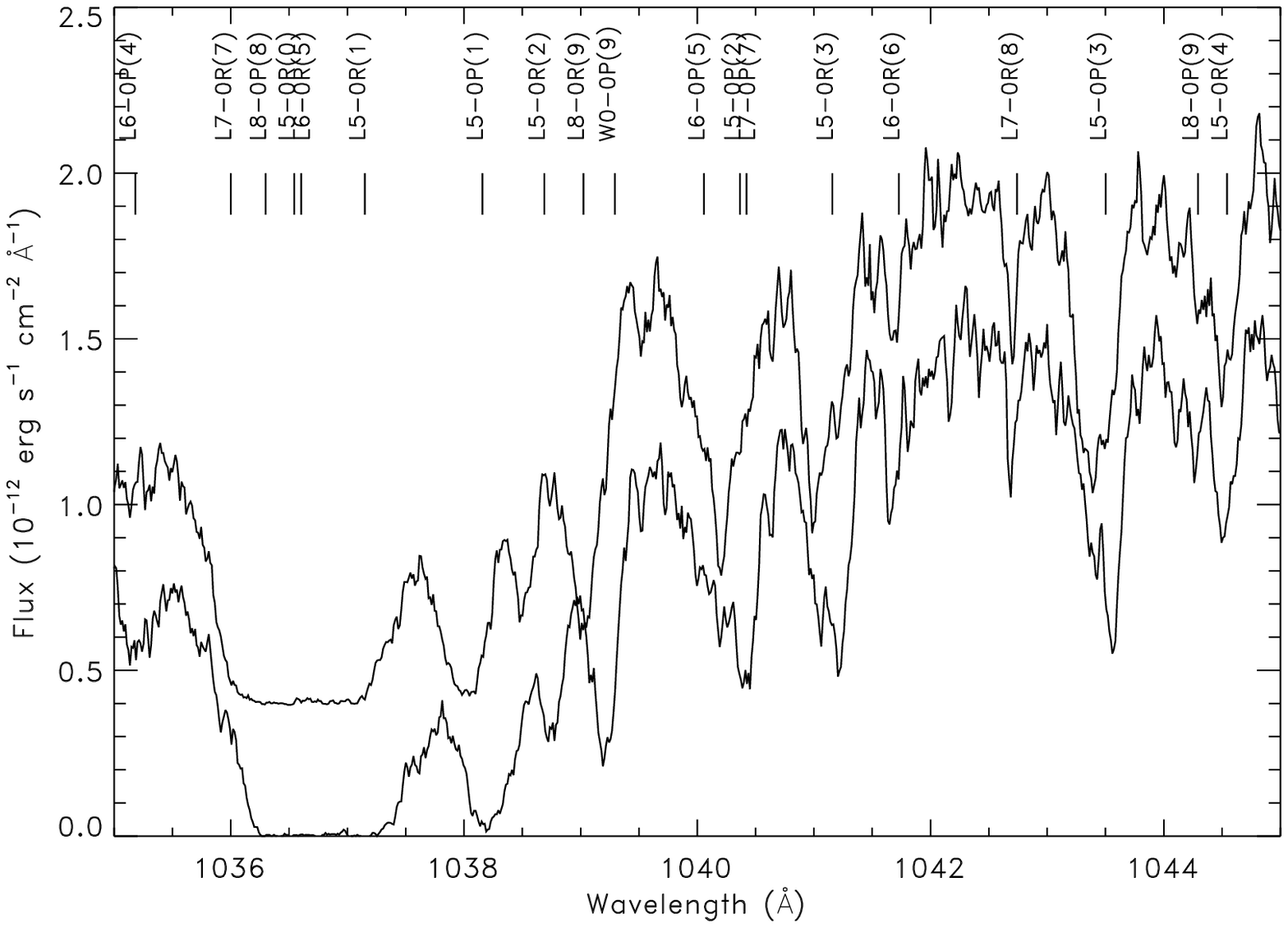}
\caption{Portion of both Herschel 36 observations; the second observation has
been shifted upward.  Note that the highly excited H$_2$ lines line up.  The
strong feature in both spectra between 1039\AA\ and 1039.5\AA\ is due to
\ion{O}{1}.}
\end{figure*}

\begin{figure*}
\plotone{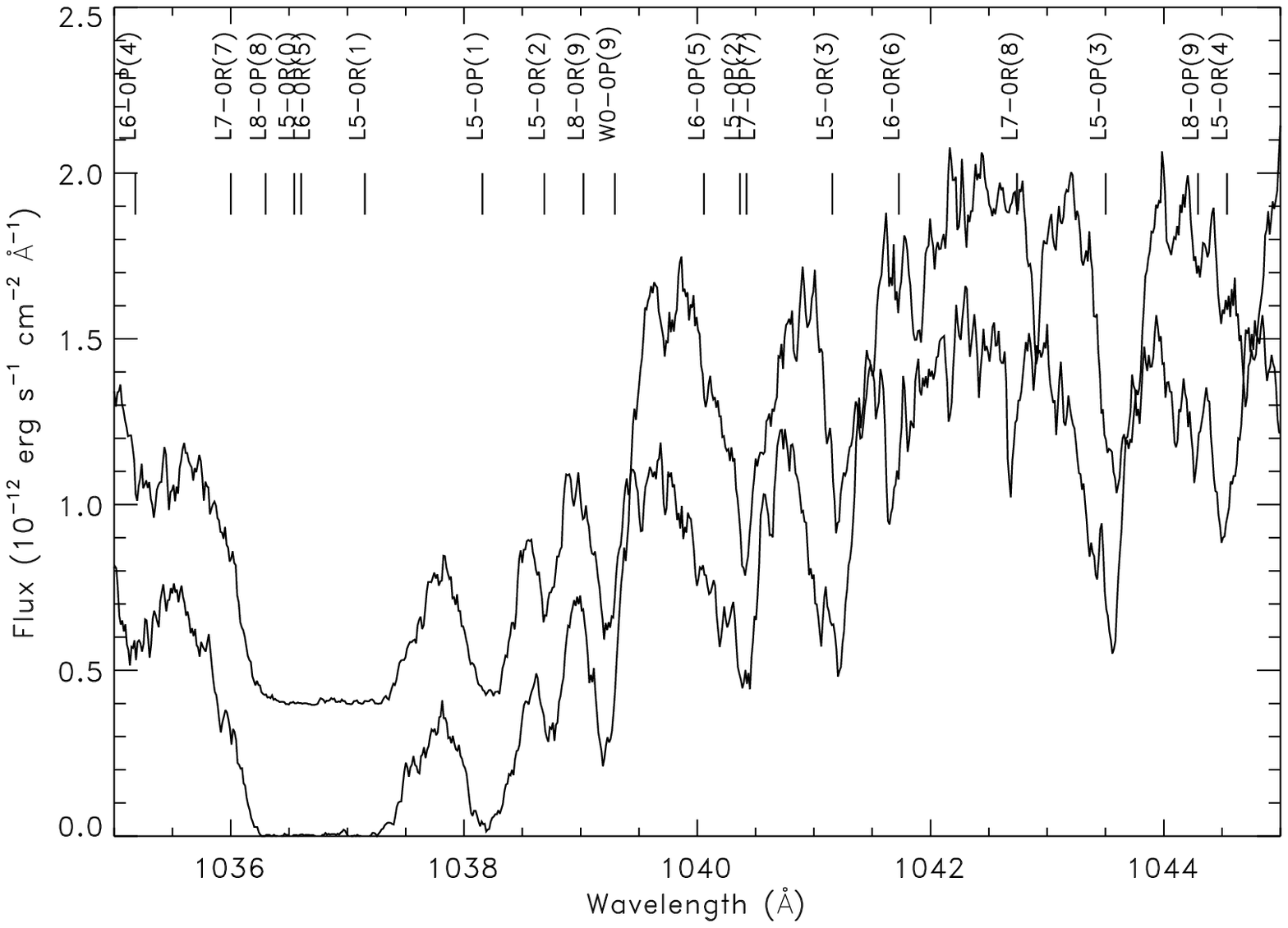}
\caption{Portion of both Herschel 36 observations; the second observation has
been shifted upward and shifted redward by 60 km s$^{1}$ to line up the H$_2$
$J$ = 0--1 bandheads.  The strong feature at 1039.2\AA\ is due to \ion{O}{1}.}
\end{figure*}

We note that the difference in the total of the geocentric and heliocentric
corrections for the two observations is 53 km s$^{-1}$; we first looked at this
data before heliocentric corrections were implemented in CALFUSE, which
resulted in a ``raw'' plot that looked like Figure 10 rather than Figure 9.
However, it seems highly unlikely that the bulk of the material -- which
appears to be in the foreground of M8 -- has undergone such a large velocity
shift.  Thus, a $\sim$60 km s$^{-1}$ redshift must have occurred in the highly
excited material relative to the bulk of the material.  This also indicates a
$\sim$60 km s$^{-1}$ deviation in the wavelength scale that was applied to the
two observations, even though it appears that the correct geocentric and
heliocentric corrections were applied.  Systematic velocity errors in Version 3
of CALFUSE can be 10 km s$^{-1}$ or more (Dixon et al 2007) in certain
circumstances.

Both {\it FUSE} observations of Herschel 36 were made through the 30$\arcsec$
square LWRS aperture.  As noted by Boiss\'{e} et al. (2009), in comparing {\it
FUSE} observations of HD 34078 taken through the LWRS aperture with an
observation taken with the 4$\arcsec$ $\times$ 20$\arcsec$ MDRS aperture, there
were subtle H$_2$ line profile/strength differences.  They attributed these
differences to the extra diffuse nebular emission passing through the factor of
900/80 larger aperture.  Based on extinction curves from Fitzpatrick \& Massa
(2007), the total extinction at 1100 \AA\ toward Herschel 36 is 8.5 mag as
compared to 6.4 mag for HD 34078, so we might expect scattered light to be a
bigger issue for Herschel 36.  Some combination of excess diffuse emission,
variations in the position of the LWRS on the sky, and variations in the path
of light through the cloud from the 3 au orbit in the triple system may
conspire to produce changes in the appearance of the two spectra of Herschel
36.  However, such an effect would have to produce the very specific large
velocity shift of the highly excited lines without large variations in the
strength or width of the lines, and simultaneously the large profile variation
specifically seen in the $J$ = 2--4 lines.  Nothing like this is seen toward HD
34078, or any other {\it FUSE} observations of which we are aware.

We are unaware of any additional strong point-like UV sources within
30$\arcsec$ of the Herschel 36 system that are visible from Earth and thus
might produce a composite spectrum.  We also note that in modeling the $J$ =
0--1 lines, adding a shifted component even with $\sim$1\% of the total
material caused the modeled profile to deviate strongly from the observed
absorption.

\subsection{Comparison with H$_2$ along other lines of sight toward M8}
Herschel 36 shows considerably more reddening than most stars in the M8 complex
(McCall et al.\ 1990).  Three stars with $E(B-V)$ = 0.30--0.45 have been
observed in the far UV at moderate-to-high resolution along lines of site
within $\sim$10 arcminutes of Herschel 36, and all are apparently associated with M8
(McCall et al.\ 1990).  Dahlstrom et al.\ (2013) give considerably more detail
on this portion of M8; here we provide the first published analysis of H$_2$
towards three stars relatively near Herschel 36.

\begin{deluxetable*}{ccccccccccc}
\tablecaption{Logarithmic H$_2$ column densities for Herschel 36 and nearby lines of sight}
\tablewidth{0pt}
\tablenum{3}
\tablehead{
  \colhead{Star} & \colhead{Ang. sep.} & \colhead{Dist.\tablenotemark{a}} &
  \colhead{E(B-V)} & \colhead{$N(H_2)$} & \colhead{$N(J=0)$} & \colhead{$N(J=1)$} &
  \colhead{$N(J=2)$} & \colhead{$N(J=3)$} & \colhead{$N(J=4)$} & \colhead{$N(J=5)$} \\
  & \colhead{(\arcmin)} & \colhead{(pc)}
}
\startdata
Herschel 36 & \phn0.00 & 0.0 & 0.87 & 20.19 & 19.92 & 19.86 &  &  &  & 16.66 \\
9 Sgr       & \phn2.95 & 1.3 & 0.35 & 20.10 & 19.85 & 19.73 & 16.60 \\
HD 164816   & \phn5.48 & 2.4 & 0.30 & 20.03 & 19.76 & 19.69 & 16.86 & 15.84 & 14.76 & 14.51 \\
HD 164906   &    10.36 & 4.5 & 0.45 & 20.23 & 19.96 & 19.89 & 16.90 & 15.89 & 14.76 & 14.29
\enddata
\tablenotetext{a}{Linear plane-of-sky separation assuming stellar distance of 1500 pc.}
\end{deluxetable*}

HD 164794 (9 Sgr) lies 3 arcminutes from Herschel 36 (1.3 pc at 1.5 kpc), and was
observed with {\it ORFEUS} in 1993.  Unfortunately, the velocity resolution was
only $\sim$90 km s$^{-1}$, insufficient to see even a broad velocity structure
in H$_2$ such as that toward Herschel 36.  HD 164816 (5.5 arcmin from Her 36) and HD
164906 (10.4 arcmin), have both been observed with {\it FUSE} at the same
resolution as the Herschel 36 observation.

The low resolution of {\it ORFEUS} limits our analysis of 9 Sgr to the $J$ =
0--2 states, and even $N$(2) is rather uncertain.  However, for these
states there is excellent agreement between 9 Sgr and HD 164816, as can be seen
in Table 3.  HD 164906 is much farther from Herschel 36, but the column densities are
still only somewhat larger than for the other lines of sight.  Importantly,
neither of the {\it FUSE} observations show any hint of the highly excited
material seen towards Herschel 36, nor the broad $J$ = 2--4 component structure, as
clearly seen in Figure 11.

\begin{figure*}
\plotone{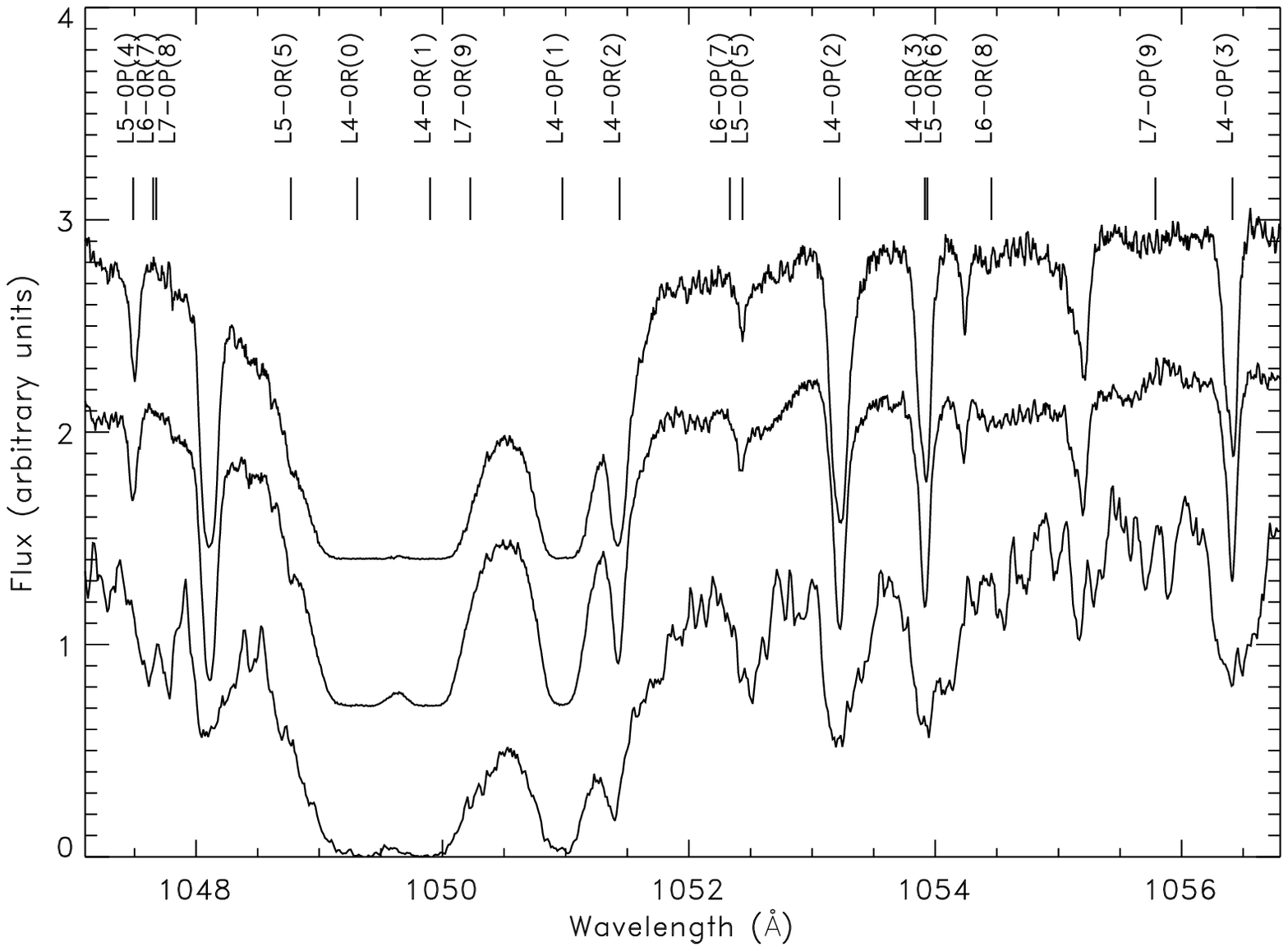}
\caption{From top to bottom; HD 164906, HD 164816, Herschel 36 second
observation.  The former spectra were normalized to roughly match the flux of
the Herschel 36 observations before being shifted upward.  The line
identifications are matched to the wavelength scale of the two HD stars, and
would be slightly redshifted to match Herschel 36.  Note the lack of velocity
structure in the $J$ = 2--4 lines towards HD 164906 and HD 164816 and the lack
of highly excited material.  The strong features at 1048 \AA\ and 1055\AA\ are
due to \ion{Ar}{1} and \ion{Fe}{2}, respectively.}
\end{figure*}

The actual H$_2$ column densities observed toward the stars in M8 are all
relatively small given the amount of extinction, i.e., $f_{\rm H2}$ $\lesssim$
0.1, but particularly for Herschel 36.  In fact, N(H$_2$) is only slightly larger
toward Herschel 36 than the other lines of sight despite having twice the reddening.
All of the stars have among the smallest molecular fractions for Galactic lines
of sight with $A_V$ $>$ 1 (e.g.\ Rachford et al.\ 2002, 2009), and Herschel 36
is {\it the} smallest of which we are aware.  The key point is that the
parameters of the bulk of the molecular material in front of a large portion of
M8 are relatively uniform and there is no evidence of unusually strong
excitation, except in the Herschel 36 observation.  Thus, the evidence points
towards the highly excited material near Herschel 36 representing the portion
of the total line of sight material that lies in a cloud (or clouds) near the
star system.  

\subsection{Molecular hydrogen and carbon monoxide emission in the vicinity of
Herschel 36}
Emission from vibrationally excited H$_2$ has been seen near Herschel 36 (Burton
2002).  The 2.12 $\mu$m 1--0 S(1) transition was detected from a roughly
bipolar shape centered on Herschel 36, with a peak flux corresponding to a $v$
= 1, $J$ = 3 column density of 1.4 $\times$ 10$^{16}$ cm$^{-2}$ uncorrected for
extinction.  Even without the correction for the IR extinction, this is still
much greater than our derived column density for this level of 1.8 $\times$
10$^{15}$ cm$^{-2}$.  In fact, Burton (2002) observed that this peak emission
occurs nearly 2 arcsec from Herschel 36.  Given the projection of the bipolar
structure on either side of Herschel 36 in the plane of the sky, little of this
material may be directly in front of the star system.  Thus, the conclusion by
Burton (2002) that the H$_2$ emission was tracing shock-heated gas does not
mean that the highly excited H$_2$ we see along the narrow beam to the stars in
the Herschel 36 system is the same material.

Similarly, the region around Herschel 36 contains one of the strongest known CO
emission sites (White et al.\ 1997).  They estimated a peak column density of
2.1 $\times$ 10$^{16}$ cm$^{-2}$ for N(C$^{\rm 18}$O) alone.  The CO peak was
located very near the star, but the beamwidths were 11--22 arcsec, depending on
the observed transition.  Since our observed N(H$_2$) appears to be much
smaller than that implied by the peak CO emission, it is not surprising that
the CO abundance observed in absorption toward Herschel 36 is quite small.
Although it is hard to precisely measure the extent of the weak rotationally
excited lines on either side of the $J$ = 0 line due to S/N issues, the E--X
0--0 band at 1076 \AA\ implies $N$(CO) $\lesssim$ 10$^{14}$ cm$^{-2}$, or a
CO/H$_2$ ratio of $\lesssim$ 6 $\times$ 10$^{-7}$, consistent with direct UV
pencil-beam derived ratios along lines of sight with similar abundances (Burgh,
et al.\ 2007, Sonnentrucker et al.\ 2007, Sheffer et al.  2008).  Clearly, both
H$_2$ and CO show considerable spatial variability in the plane of the sky,
and/or there is considerable material behind Herschel 36.  This is consistent
with the complex environment of Herschel 36 summarized by Dahlstrom et al.\
(2013).

\section{Conclusions and future directions}
The region near Herschel 36 is very complex and has been studied at a variety
of wavelengths.  Our far-UV observations provide the first look at H$_2$ along
the line of sight towards the UV-bright Herschel 36 star system.

We have found an absorption component displaying highly excited H$_2$
apparently lying near the Herschel 36 system which is subjected to an intense
radiation field and experienced a $\sim$60 km s$^{-1}$ redshift during a
3.6-year period.  This component does not appear to be at the same velocity as
the material which dominates the lower excitation, and this latter material
does not appear to produce visible absorption beyond $J$ = 5.  Given the width
of the $J$ = 2--3 lines in the spectrum, multiple additional components may
exist and/or there may be material in a relative narrow physical distance range
but spread across a large velocity range.  The highly excited component -- the
primary focus of this paper -- appears to only contain a small fraction of the
total H$_2$ along the line of sight on the order of 1\% . 

While certain portions of the material surrounding the Herschel 36 system may
be associated with outflows or jets and thus be shock heated (Burton 2002), the
material in the immediate foreground seen in our observations appears similar
to that seen in front of HD 37903 which has been interpreted as the result of
fluorescent excitation due to the strong UV field near the stars (Meyer et al.\
2001; Gnaci\'{n}ski 2011).

Other unusual excitation has been seen toward the Herschel 36 system.
Dahlstrom et al.\ (2013) reported the first known detection of rotationally
excited CH and CH$^{+}$ in absorption.  This excitation is attributed by Oka et
al. (2013) to far-IR radiation from a source $\sim$375 au from the main stellar
triplet, believed to be a heavily embedded B-type star (Goto et al.\ 2006).
Unusual redward wings have been seen in some Diffuse Interstellar Bands (DIBs)
toward Herschel 36, again thought to trace high levels of rotational excitation
of the molecule(s) responsible for the DIBs (Dahlstrom et al.\ 2013).  The
environment near the Herschel 36 system is thus an important laboratory for
studying the effects of the intense radiation fields in young star clusters on
gas and dust.

{\it HST} observations with {\it STIS} would provide a wealth of additional
information on Herschel 36.  First, we could much more precisely study the
highly excited H$_2$ via the hundreds of ro-vibrational transitions available
longward of 1150 \AA\ .  This would permit precise enough column densities for
a detailed modeling analysis of the excitation.  Second, the same observations
will provide high resolution observations of numerous atomic and molecular
species.  This will allow us to assess the true velocity structure down to 3 km
s$^{-1}$ resolution, and potential information on the physical conditions in
the various components.

\acknowledgments
This work is based on data obtained for the Guaranteed Time Team by the
NASA-CNES-CSA {\it FUSE} mission operated by the Johns Hopkins University.
This research has made use of the SIMBAD database, operated at CDS, Strasbourg,
France.  We wish to thank D. York, D. Welty, L. Hobbs, and the anonymous
referee for helpful comments.  BLR would like to thank the University of
Colorado's Center for Astrophysics and Space Astronomy for use of facilities
during a sabbatical leave when much of this work was completed.  Partial
support for this work has been provided by the National Science Foundation,
under grant AST-1008801 (BLR).

{\it Facilities:} \facility{FUSE}, \facility{ORFEUS}

\end{document}